\documentclass[aps,prl,twocolumn,showpacs,superscriptaddress,reprint,
nofootinbib,tightenlines,preprintnumbers]{revtex4-1}

\usepackage[utf8]{inputenc}
\usepackage[english]{babel}
\usepackage{amssymb,amsthm,amsmath,amstext,amsbsy,amsopn}
\usepackage{bbm}
\usepackage{nicefrac}
\usepackage{slashed}
\usepackage{pstricks}
\usepackage{graphicx}
\usepackage{hyperref}
\usepackage{leftidx}
\usepackage{environ}
\usepackage{mathtools}
\usepackage{xspace}
\usepackage{suffix}

\newcommand{\ie}{\textit{i.e.}\xspace}
\newcommand{\eg}{\textit{e.g.}\xspace}

\newcommand{\apriori}{\textit{a priori}\xspace}
\WithSuffix\newcommand\apriori*{\textit{a-priori}\xspace}

\newcommand{\etc}{\textit{etc.}\xspace}

\newcommand{\mathspace}{\ \ }
\newcommand{\mathtext}[1]{\mathspace\text{#1}\mathspace}

\newcommand{\fm}{\ensuremath{\mathrm{fm}}}

\newcommand{\dd}{\mathrm{d}}

\newcommand{\gsim}{\gtrsim}

\newcommand{\ii}{\mathrm{i}}

\newcommand{\vcalD}{\boldsymbol{\cal D}}

\newcommand{\OO}{\mathcal{O}}

\newcommand{\eps}{\varepsilon}

\newcommand{\bra}[1]{\langle #1|}

\newcommand{\ket}[1]{|#1\rangle}

\newcommand{\braket}[2]{\langle #1|#2\rangle}

\newcommand{\MN}{M_N}

\newcommand{\Mpi}{M_\pi}

\newcommand{\ThreeSOne}{\ensuremath{{}^3S_1}\xspace}
\newcommand{\OneSNot}{\ensuremath{{}^1S_0}\xspace}

\newcommand{\Triton}{\ensuremath{{}^3\mathrm{H}}\xspace}
\newcommand{\ThreeH}{\Triton}
\newcommand{\ThreeHe}{\ensuremath{{}^3\mathrm{He}}\xspace}
\newcommand{\FourHe}{\ensuremath{{}^4\mathrm{He}}\xspace}

\newcommand{\SixLi}{\ensuremath{{}^6\mathrm{Li}}\xspace}

\newcommand*\rvec[1]%
{\ensuremath{\overset{\smash{\raisebox{-1.5pt}{\tiny$\rightarrow$}}}{#1}}}
\newcommand*\lvec[1]%
{\ensuremath{\overset{\smash{\raisebox{-1.5pt}{\tiny$\leftarrow$}}}{#1}}}

\newcommand{\vp}{\mathbf{p}}

\newcommand{\MeV}{\ensuremath{\mathrm{MeV}}}

\newcommand{\sti}{\mathbf{i}}

\newcommand{\LO}{\text{LO}\xspace}
\newcommand{\NLO}{\text{NLO}\xspace}
\newcommand{\NNLO}{\text{N$^2$LO}\xspace}

\newcommand{\lmax}{\ensuremath{l_{\text{max}}}}
\newcommand{\jmax}{\ensuremath{j_{\text{max}}}}

\NewEnviron{subalign}[1][]{%
\begin{subequations}\begin{align}
  \BODY
\end{align}\label{#1}\end{subequations}
}

\NewEnviron{spliteq}{%
\begin{equation}\begin{split}
  \BODY
\end{split}\end{equation}
}

\begin{document}

\title{Nuclear Physics Around the Unitarity Limit}

\author{Sebastian König}
\email{koenig.389@osu.edu}
\affiliation{Department of Physics, The Ohio State University,
Columbus, Ohio 43210, USA}
\affiliation{Institut für Kernphysik, Technische Universität Darmstadt, 
64289 Darmstadt, Germany}
\affiliation{ExtreMe Matter Institute EMMI,
GSI Helmholtzzentrum für Schwerionenforschung GmbH,
64291 Darmstadt, Germany}

\author{Harald W. Grießhammer}
\email{hgrie@gwu.edu}
\affiliation{Institute for Nuclear Studies, Department of 
Physics, George Washington University, Washington DC 20052, USA}

\author{H.-W. Hammer}
\email{Hans-Werner.Hammer@physik.tu-darmstadt.de}
\affiliation{Institut für Kernphysik, Technische Universität Darmstadt, 
64289 Darmstadt, Germany}
\affiliation{ExtreMe Matter Institute EMMI,
GSI Helmholtzzentrum für Schwerionenforschung GmbH,
64291 Darmstadt, Germany}

\author{U. van Kolck}
\email{vankolck@ipno.in2p3.fr}
\affiliation{Institut de Physique Nucléaire, CNRS-IN2P3, 
Univ. Paris-Sud, Université Paris-Saclay, 91406 Orsay, France}
\affiliation{Department of Physics, University of Arizona,
Tucson, AZ 85721, USA}

\date{\today}

\begin{abstract}
  We argue that many features of the structure of nuclei emerge from a
  strictly perturbative expansion around the unitarity limit, where the
  two-nucleon $S$ waves have bound states at zero energy.  In this limit, the
  gross features of states in the nuclear chart are correlated to only
  one dimensionful parameter, which is related to the breaking of scale
  invariance to a discrete scaling symmetry and set by the triton binding
  energy.  Observables are moved to their physical values by small,
  \emph{perturbative} corrections, much like in descriptions of the fine
  structure of atomic spectra.  We provide evidence in favor of the conjecture 
  that light, and possibly heavier, nuclei are bound weakly enough to be 
  insensitive to the details of the interactions but strongly enough to be 
  insensitive to the exact size of the two-nucleon system.
\end{abstract}

\maketitle

For the purposes of nuclear physics, QCD, the theory of strong interactions,
has essentially two independent parameters, namely the up and down quark
masses.  Their average controls the pion mass and consequently the range of
the nuclear force, $R\sim \Mpi^{-1} \simeq 1.4~\fm$.  Their difference, plus
electromagnetism, generates small differences in masses and interactions
between neutrons and protons.  At the physical point, the two-nucleon ($N\!N$) 
scattering length in the \ThreeSOne channel is $a_t\simeq 5.4~\fm$, with the 
deuteron as a shallow bound state ($B_D\simeq 2.224$ MeV); in the \OneSNot 
channel, $a_s\simeq {-}23.7~\fm$, and a shallow virtual bound state exists at
$B_{N\!N^*}\simeq 0.068$ MeV.  
With relatively small changes in quark masses, these states become, 
respectively, unbound
and bound~\cite{Beane:2001bc,Beane:2002vs,Epelbaum:2002gb,Beane:2002xf}. 
In the physics of cold atoms near Feshbach resonances, external magnetic fields 
play a role similar to the quark masses and allow the scattering length to be 
tuned arbitrarily~\cite{Braaten:2004rn}.

Approximate correlations $B_{D,N\!N^*}\approx1/(\MN a_{t,s}^2)$, with 
$\MN\simeq 940$ MeV the nucleon mass, hold because the size of all these scales 
is unnatural compared to the typical interaction range $R$.  The $N\!N$ system 
therefore appears close to the unitarity (or unitary) limit, where both states 
cross zero energy, the scattering lengths become infinite ($1/a_{t,s}=0$), and 
cross sections saturate the unitarity bound.  It has indeed been suggested that 
this happens not far from the physical point~\cite{Braaten:2003eu}.  While this 
presumed proximity has been discussed qualitatively for a long time, it has 
traditionally not played any special role in constructing nuclear forces, 
and it is neither assessed nor exploited in order to simplify the description 
of nuclei.  As an exception, Refs.~\cite{Kievsky:2015zga,Kievsky:2015dtk} use 
potential models to map out correlations between observables in three- and 
four-nucleon systems as the limit is approached at fixed $a_t/a_s$.

Here we argue that the typical particle binding momentum $Q_A$ of the 
$A$-nucleon system satisfies $1/a_{s,t} < Q_A < 1/R$, so that a combined
expansion in $Q_AR$ and $1/(Q_Aa_{s,t})$ converges quickly and 
\emph{quantitatively} reproduces the physical systems.  With this, the gross 
features of states in the nuclear chart are determined by a very simple 
leading-order interaction (governed by a single parameter), whereas, much like 
the fine structure of atomic spectra, observables are moved to their physical 
values by small \emph{perturbative} corrections.  Our conjecture places nuclei 
in a sweet spot: bound weakly enough to be insensitive to the details of the 
interaction, but dense enough to be insensitive to the exact values of the 
large two-particle scattering lengths.  One might surmise that due to the 
absence of scales, a theory at the unitarity limit allows only for trivial 
observables, like bound states with zero or infinite energies.  However,
the nontrivial renormalization of the three-body system introduces instead 
exactly one new dimensionful parameter, which sets the scale for all few-body 
observables.  Indeed, the energies of bosonic clusters near unitarity
are determined in terms of the trimer 
energy~\cite{vonStecher:2010,Bazak:2016wxm}.

In the following, we provide explicit evidence that our conjecture holds for the
binding energies of three and four nucleons.  Since the $N\!N$ binding energies
are small, their dynamics is dominated by large distances and small momenta,
the regime of the effective range expansion (ERE)~\cite{Bethe:1949yr}.  Its
consequences are captured by an effective field theory (EFT) which, apart from 
long-range electromagnetic interactions mediated by photons ($A_\mu$),
contains only contact interactions between nonrelativistic nucleon isospin
doublets $N=(p\;n)^T$ of proton and neutron fields.  Following the notation of
Ref.~\cite{Konig:2015aka}, its Lagrange density is
\begin{multline}
 \mathcal{L} = N^\dagger\left(\ii {\cal D}_0+\frac{\vcalD^2}{2\MN}\right)N \\
 \null + \sum\nolimits_{\sti}C_{0,\sti}
 \left(N^T P_{\sti} N\right)^\dagger \left(N^T P_{\sti} N\right)
 + D_0 \left(N^\dagger N\right)^3 +\cdots \,,
\label{eq:L-Nd}
\end{multline}
where ${\cal D}_\mu = \partial_\mu + \ii eA_\mu (1+\tau_3)/2$, $e$ is the proton 
charge, $\tau_a$ a Pauli matrix in isospin space, and $P_{\sti}$ projectors onto 
the $N\!N$ $S$ waves.  $C_{0,\sti}$, $D_{0}$, \etc are ``low-energy constants'' 
(LECs) determined from QCD or experiment.  This ``Pionless EFT''reproduces the 
ERE in the $N\!N$ sector~\cite{vanKolck:1997ut,Kaplan:1998tg,Kaplan:1998we,
vanKolck:1998bw,Chen:1999tn} but extends it to an arbitrary number of particles
and interactions with external fields.  The two-body  interactions with LECs 
$C_{0,\sti}$ are related to $a_{s,t}$, while higher-derivative interactions are 
associated with the effective ranges and higher ERE parameters, as well as 
higher partial waves.  The organizational principle (``power counting'') 
attributes the $C_{0,\sti}$ to nonperturbative leading-order (\LO), and 
higher-derivative interactions to subleading orders.  These are added 
perturbatively and include the effects of the interaction range $R$ in a 
systematic expansion in $QR\ll1$, where $Q$ is a typical low-momentum scale.

Stability of light nuclei results from an additional LO interaction, a single
nonderivative three-nucleon ($3N$) contact
interaction~\cite{Bedaque:1998kg,Bedaque:1998km,Bedaque:1999ve}, with LEC
$D_0$.  Derivative corrections to this $3N$ interaction start at
next-to-next-to-leading order
(\NNLO)~\cite{Bedaque:1998km,Bedaque:2002yg,Ji:2012nj,Vanasse:2013sda,
Griesshammer:2005ga}.  Little is known about the orders at which higher-body
interactions appear, except that they are not LO~\cite{Platter:2004he,
Platter:2004zs,Platter:2005,Hammer:2006ct,Kirscher:2009aj,Kirscher:2011uc,
Barnea:2013uqa,Kirscher:2015ana,Kirscher:2015yda,Bazak:2016wxm,
Contessi:2017rww}.  Based on a zero-range model,
Refs.~\cite{Yamashita:2006,Hadizadeh:2011qj,Hadizadeh:2011ad} report some
sensitivity of four-body energies to a four-body scale, but these results do
not contradict the absence of a four-body interaction at LO in Pionless EFT.
The absence of an essential four-body parameter has also been verified in the
context of potential models with short range~\cite{Hanna:2006,
vonStecher:2009,Deltuva:2010xd,Gattobigio:2012tk} and of renormalization-group
analyses~\cite{Schmidt:2009kq,Avila:2013rda,Horinouchi:2014ata}.  As a
consequence, the $3N$ LO strength parameter $\Lambda_\star$, together with the
LO two-body interactions, determines the spectrum and scattering for systems
with more particles~\cite{Platter:2004he,Platter:2004zs,Platter:2005,
Hammer:2006ct,Stetcu:2006ey,Kirscher:2009aj,Kirscher:2011uc,Barnea:2013uqa,
Kirscher:2015ana,Kirscher:2015yda,Lensky:2016djr,Bazak:2016wxm,
Contessi:2017rww}.  This single relevant $3N$ parameter generates
correlations among few-body observables such as the
Phillips~\cite{Phillips:1968zze} and Tjon~\cite{Tjon:1975sme} lines.

The standard pionless formulation with finite scattering lengths as LO input
explicitly breaks two important symmetries: first, the $SU(4)_W$ Wigner
symmetry of combined spin and isospin transformations~\cite{Wigner:1936dx} is
broken in the two-body sector for $a_t \neq a_s$~\cite{Mehen:1999qs}, while it
is obeyed by the $3N$ interaction~\cite{Bedaque:1999ve}.  Second, discrete
scale invariance leads to the log-periodic shape of the running coupling
$D_0$~\cite{Bedaque:1998kg,Bedaque:1998km,Bedaque:1999ve} and to an infinite
geometric tower of Efimov states in the three-body system~\cite{Efimov:1970zz},
both determined by the $3N$ LO strength parameter $\Lambda_\star$
(see Eq.~\eqref{eq:limcyc} below).  It, too, is broken for 
$a_{s,t}\neq\infty$. (Note that in contrast to
Refs.~\cite{Efimov:1970zz,Efimov:1979zz,Braaten:2004rn}, scattering lengths
are not rescaled with the discrete scaling factor here.)

The unitarity limit manifestly respects both symmetries and has 
$\Lambda_\star$ left as single parameter.  In our calculation, 
this is fixed at \LO to reproduce the physical triton (degenerate with \ThreeHe 
at this order) as one of the Efimov states.  In fact, the $3N$ and $4N$ systems 
in the unitarity limit decouple into a symmetric piece, identical to a 
formulation of three- and four-boson systems, and an antisymmetric piece.  At 
unitarity, a three-boson Efimov state with binding energy $B_3$ is associated 
with two four-boson states~\cite{Hammer:2006ct}---one relatively deep at 
$B_4/B_3\simeq 4.611$, and one barely below the particle-trimer threshold, 
$B_{4^*}/B_3\simeq 1.002$~\cite{Deltuva:2010xd}.  For the $\alpha$ particle, 
the ground state is at $B_\alpha/B_H\simeq 3.66$ and the excited state at 
$B_{\alpha^*}/B_H\simeq 1.05$, where $B_H\simeq 7.72~\MeV$ is the \ThreeHe binding 
energy.  In addition, models (see \eg Ref.~\cite{Adhikari:1982zz}) indicate the 
existence of a virtual $3N$ state at $B_{T^*} \lesssim0.5~\MeV$ for physical 
scattering lengths, which becomes the second Efimov state as $B_D$ is decreased. 
Thus, the $3N$ and $4N$ spectra are consistent with mildly broken discrete 
scale invariance, suggesting a perturbative treatment, described below.

For the calculation, the LO two-body potentials derived from 
Eq.~\eqref{eq:L-Nd} are written as
\begin{equation}
 V_2^{(0)} = \sum\nolimits_{\sti} C_{0,\sti}^{(0)} \, 
 \ket{\sti}\ket{g}\bra{g}\bra{\sti} \,,
\label{eq:V2}
\end{equation}
where $\ket{\sti}$ collects the spin-isospin structure and $\ket{g}$
implements a separable regularization.  With $\vp$ the momentum corresponding
to the kinetic energy $E$ in the $N\!N$ center-of-mass system, $g(p)\equiv
\braket{\vp}{g}$ satisfies $g(0)=1$ and $g(p\gg \Lambda)\ll 1$ for arbitrary
cutoff $\Lambda$.  The results shown below have been obtained with two different
implementations of the theory.  For the \ThreeHe calculation, we follow
Ref.~\cite{Konig:2015aka}, which uses a sharp cutoff (step function) regulator
$g_{s}(p)= \theta (\Lambda-p)$ and includes the two-body interactions through
dibaryon auxiliary fields, in lieu of Eq.~\eqref{eq:V2}.  Our new 
Faddeev(-Yakubovsky) calculations use a convenient Gaussian regulator instead,
$g_{G}(p) =\exp(-p^2/\Lambda^2)$.  Since only the \LO calculation mandates a
nonperturbative treatment, we use (distorted-wave) perturbation theory for
higher orders, \ie, \NLO results depend only linearly on \NLO contributions.
These are:
\textit{(i)} terms as in Eq.~\eqref{eq:V2} accounting for LEC shifts
$C_{0,\sti}^{(0)} \to C_{0,\sti}^{(1)}$, corresponding to the expansion
$C_{0,\sti} = C_{0,\sti}^{(0)} + C_{0,\sti}^{(1)} + \cdots$ and shifting
to the physical values of $a_{s,t}$;
\textit{(ii)} one Coulomb interaction $\sim e^2\MN/(4\pi Q)$ in the $pp$ 
channel;
\textit{(iii)} one isospin-breaking contact interaction in the $pp$ channel as
in Eq.~\eqref{eq:V2} with $C_{0,s}^{(0)} \to \Delta
C_{0,s,pp}^{(1)}$, required for proper renormalization of Coulomb
effects;
and \textit{(iv)} one range correction per $N\!N$ $S$ wave.

In order to treat Coulomb effects in the $3N$ sector perturbatively,
Ref.~\cite{Konig:2015aka} expanded the $\OneSNot$ channel around the unitary
limit.  Here we also expand the $\ThreeSOne$ channel in $1/(Q_3a_t)$, which
is a significantly more radical simplification, given that $a_t$ is not nearly
as large as $a_s$.  The two-body amplitude is a geometric series that can be
resummed analytically for a separable regulator.  We remove the arbitrary
$\Lambda$ dependence from observables by demanding that the two inverse
scattering lengths vanish at \LO and enter linearly at \NLO.  Renormalization
is achieved if
\begin{equation}
 C_{0,\sti}^{(0)} = \frac{{-}2\pi^2}{\MN\Lambda}\theta^{-1}
 \mathtext{,}
 C_{0,\sti}^{(1)} = \frac{\MN}{4\pi a_\sti} C_{0,\sti}^{(0)2} \,,
\label{eq:C00-C01}
\end{equation}
where $\theta=\int\dd q\,g^2(q)/\Lambda$ is a regulator-dependent pure number.
The LO amplitude takes then the scale invariant form $T_\sti^{(0)}(E) \propto
1/\sqrt{\mathstrut{-}\MN E - \ii\eps}$, with \NLO corrections proportional to
$C_{0,\sti}^{(1)}$.  The deuteron binding energy vanishes up to \NLO, but
$(B_D)^\NNLO = 1/(\MN a_t^2)\simeq 1.41~\MeV$ coincides with the standard 
zero-range value.  For more details, see Ref.~\cite{Konig:2016iny}.

A $3N$ potential is needed at LO for renormalizability, \ie, to avoid the
Thomas collapse~\cite{Thomas:1935zz} and ensure that three-body observables
have a well-defined limit for $\Lambda\gg
1/R$~\cite{Bedaque:1998kg,Bedaque:1998km,Bedaque:1999ve}.  We choose a
separable form
\begin{equation}
 V_3^{(0)} = D_0^{(0)} \, \ket{\Triton}\ket{\xi}\bra{\xi}\bra{\Triton} \,,
\label{eq:V3}
\end{equation}
where $\ket{\Triton}$ denotes the projector onto the $J=S=T=\nicefrac12$ $3N$
state and $\ket{\xi}$ the regulator, either sharp or
$\braket{u_1 u_2}{\xi}=g_{G}(\sqrt{\mathstrut u_1^2+3u_2^2/4})$ for Jacobi 
momenta $u_{1,2}$.  We take the triton binding energy as the one observable 
needed to fix $D_0^{(0)}(\Lambda)$.

An \apriori estimate of the typical $A$-body scale equidistributes the
total binding energy amongst its constituents, \ie~$Q_A\sim\sqrt{2M_NB_A/A}$.  
For three nucleons, $Q_3\approx70\;\MeV$ appears indeed in the sweet spot, 
namely larger than $1/a_{s,t}\lesssim45\;\MeV$ but smaller than the breakdown 
scale of Pionless EFT (expected to be about $140\;\MeV$).  We thus compare the
running of $D_0^{(0)}$ at full-unitarity LO with the result of standard 
pionless LO (scattering lengths at their physical values) and of $\OneSNot$
unitarity~\cite{Konig:2015aka}.  Taking the same sharp-momentum regulator, 
all three cases agree well with
the log-periodic form~\cite{Bedaque:1998kg,Bedaque:1998km,Bedaque:1999ve}
\begin{equation}
 D^{(0)}(\Lambda) \propto\frac{1}{\Lambda^4}
 \frac{\sin \left(s_0 \log (\Lambda/\Lambda_*)-\arctan s_0^{-1}\right)}
 {\sin \left(s_0 \log (\Lambda/\Lambda_*)+\arctan s_0^{-1}\right)} \,,
\label{eq:limcyc}
\end{equation}
where $s_0\simeq 1.00624$.  The proportionality factor is scheme and regulator
dependent.  We find $\Lambda_* = 175$, $168$, $234~\MeV$ for standard Pionless
EFT, $\OneSNot$ unitarity, and full unitarity, respectively.  The changes at
$\OneSNot$ and full unitarity go in opposite directions since the \OneSNot
interaction is more attractive at unitarity than for the physical scattering
length as LO input, while the \ThreeSOne interaction is less attractive.

\begin{figure}[tb]
\centering
\includegraphics[width=0.45\textwidth,clip]{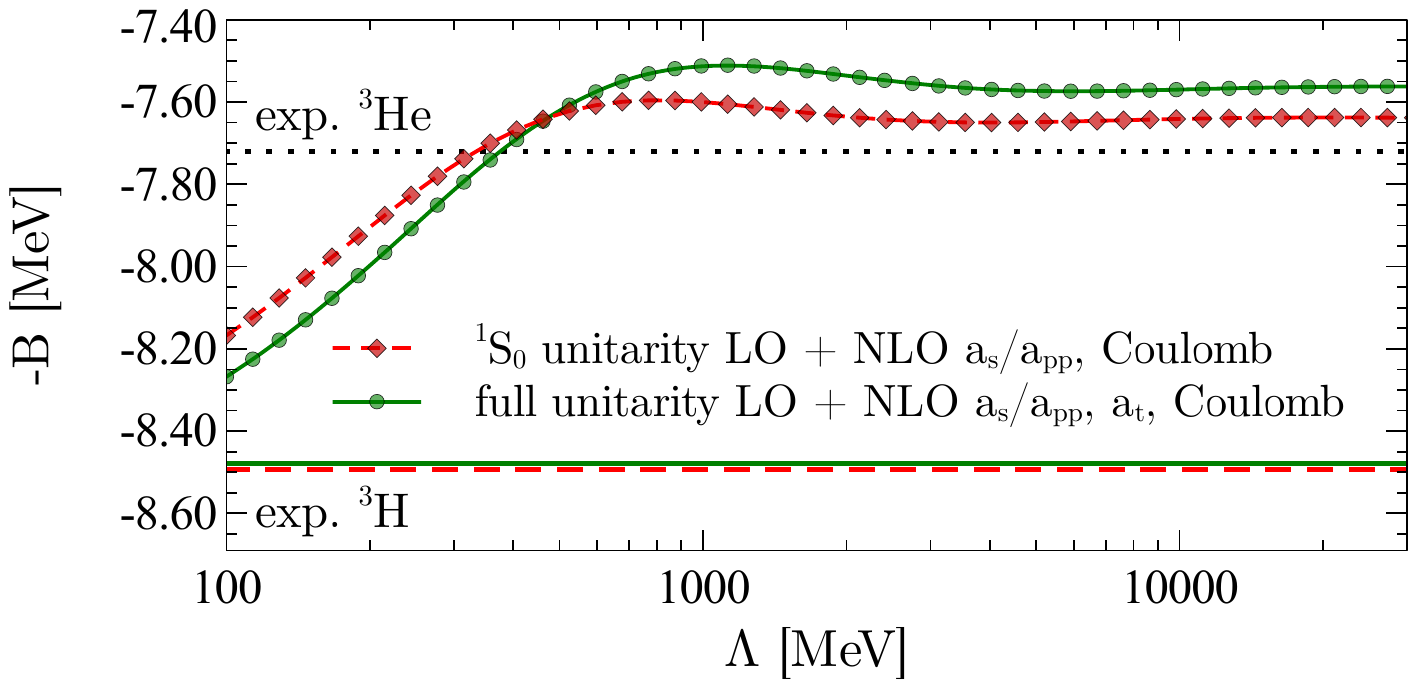}
\caption{Three-nucleon binding energies at \NLO as function of
  the $3N$ sharp cutoff. (Red) dashed curves: results of
  Ref.~\cite{Konig:2015aka}, keeping the physical $a_{t}$ at LO.  (Green) 
  solid curves: effect of taking both $S$ waves to the unitarity limit at \LO 
  and then including the physical values perturbatively at \NLO, along with 
  one-photon exchange and its counterterm.  Horizontal lines: experimental 
  values. Top (bottom) lines: \ThreeHe (\ThreeH, fitted).}
\label{fig:En-3He-UU}
\end{figure}

At \NLO the $3N$ interaction has the same form as in Eq.~\eqref{eq:V3}, with 
$D_0^{(0)} \to D_0^{(1)}$ chosen to keep the triton energy unchanged.  
As a nontrivial check, we repeat the \ThreeHe calculation of
Ref.~\cite{Konig:2015aka} with a full-unitarity LO and add finite 
$1/a_{s,t}$ plus one-photon exchange and its counterterm as \NLO corrections.
Figure~\ref{fig:En-3He-UU} shows excellent agreement, with only small
differences to $\OneSNot$ unitarity.  Convergence with the cutoff is evident.  
We predict a triton-helion splitting $(B_T-B_H)^\NLO \simeq (0.92 \pm 
0.18)~\MeV$, compared to the experimental value $(B_T-B_H)^{\rm exp}\simeq 
0.764~\MeV$.  Our $20\%$ error estimate follows Ref.~\cite{Konig:2015aka} and is 
larger than the additional $\OO(1/(Q_3 a_t)^2)$ from the new expansion.

At full unitarity, the LO spectrum consists of a series of states spaced by a
factor of $\exp(2\pi/s_0)\approx515$~\cite{Efimov:1970zz}.  An infinite number
of states shallower than the triton/helion accumulates at zero energy.  These
lie outside the range of applicability of our expansion since their typical 
momenta are not large compared to $1/|a_{s,t}|$.  Nevertheless looking at 
their perturbative shifts, we find that at \NLO they remain an order of 
magnitude shallower than the \NNLO deuteron, and generally we expect them to 
disappear above threshold at \NNLO.  In addition, with increasing cutoff, deeper 
$3N$ states enter the spectrum with binding momenta well above $1/R$.  These are 
well outside the range of validity of the EFT and thus not a fundamental issue, 
but they complicate the numerical solution of the $4N$ problem.  For now, we 
restrict our $4N$ calculations to a $\Lambda$ range in which these are absent.  
For simplicity, we neglect electromagnetic and range corrections at \NLO, 
focusing on the $1/(Q_4a_t)$ expansion.

Our $4N$ calculation follows
Refs.~\cite{Platter:2004he,Platter:2004zs,Platter:2005} (based, in turn, on
Ref.~\cite{Kamada:1992aa}) with an independently developed numerical
implementation.  We include a sufficient number of angular components to ensure 
numerical convergence; see a subsequent publication for
details~\cite{Koenig:2016xx}. 

\begin{figure}[tb]
\centering
\includegraphics[width=0.45\textwidth,clip]{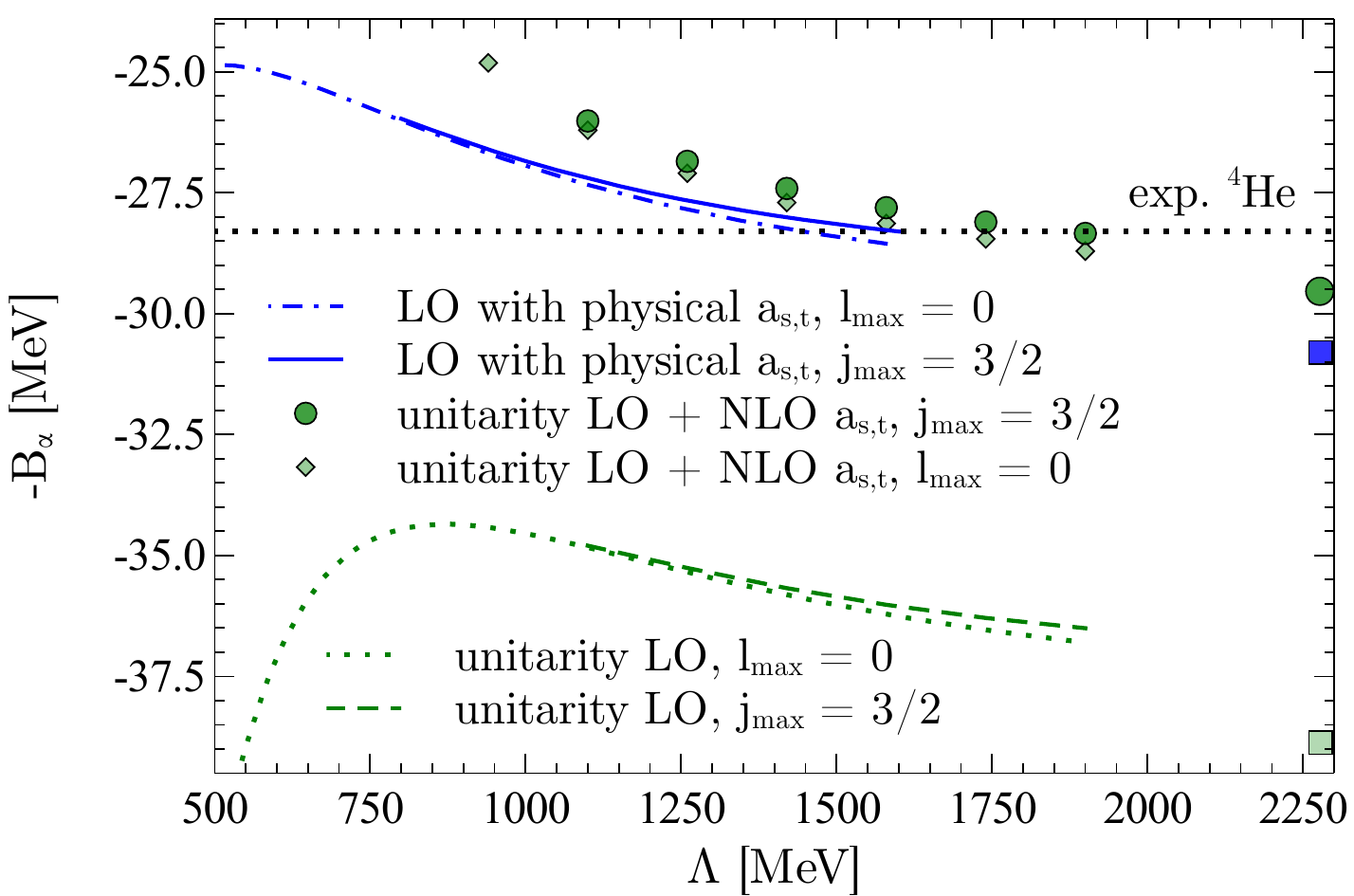}
\caption{\FourHe binding energy as function of the Gaussian
  cutoff.  (Blue) solid and (green) dashed lines: standard Pionless EFT and
  full unitarity at LO, respectively, with partial waves up to
  $\jmax=\nicefrac32$.  (Blue) dot-dashed and (green) dotted line: same for
  $S$ waves only ($\lmax=0$).  
  (Green) diamonds ($\lmax=0$) and circles ($\jmax=\nicefrac32$):
  first-order corrections in $1/a_{s,t}$ are added.
  Results for $\jmax = 2$ are almost identical to $\jmax=\nicefrac32$ and
  not shown.  Large symbols on right edge: $\Lambda \to \infty$
  extrapolation (see text).}
\label{fig:En-4He-UU-atad}
\end{figure}

Figure~\ref{fig:En-4He-UU-atad} shows results for the $\alpha$-particle
binding energy $B_\alpha(\Lambda)$. They are well fitted with a quadratic
polynomial in $1/\Lambda$ for large $\Lambda$, which we therefore use to
extrapolate $\Lambda\to \infty$. For standard Pionless EFT at LO, they are
consistent with Refs.~\cite{Platter:2004he,Platter:2004zs,Platter:2005}.
Implementing the unitarity-limit at LO leads to about $10$ MeV more binding,
as expected from a more attractive $3N$ interaction.  We find a bound excited 
state just below the proton-triton breakup threshold, in contrast to $0.3~\MeV$ 
above as indicated by experiment. The LO results $(B_\alpha/B_T)^\text{LO}=4.66$
for the ground and $(B_{\alpha^*}/B_T)^\text{LO}=1.002$ for the excited state
of \FourHe agree with four-boson
unitarity~\cite{Platter:2005,Hammer:2006ct,Deltuva:2010xd}. 

Remarkably, the first $1/(Q_4a_t)$ correction brings us very close to the
standard Pionless EFT result.  That is perhaps accidental due to the highly
symmetrical nature of the $\alpha$ particle, the level of agreement in the
helion being perhaps more representative.  All results in
Fig.~\ref{fig:En-4He-UU-atad} are uncorrected for electromagnetic and range
effects.  At present, no calculation of these effects in Pionless EFT exist,
except for Ref.~\cite{Lensky:2016djr} where higher orders were however not
treated perturbatively.  With the uncertainty expected to be dominated by
range corrections, $\OO(r_{s,t}/a_{s,t}) \simeq 30\%$, we obtain
$(B_\alpha)^{\NLO (r=0)} = 29.5\pm 8.7$ MeV with zero effective ranges.  The
ratio $(B_\alpha/B_T)^{\NLO (r=0)}\approx3.48$ is in good agreement with
$(B_\alpha/B_T)^\text{exp}=3.34$.

While slow convergence for the excited-state wavefunction puts its full \NLO 
calculation beyond our current computational resources, we used a four-boson 
model tuned such that its \LO is similar to the \FourHe system, namely choosing 
$B_3=8.5~\MeV$.  Since the four-boson and $4N$ systems are identical for exact 
$SU(4)_W$ symmetry, this is an adequate rendering of the more complex physical 
world.  Calculations with $a_2=20~\fm$ and $a_2 = 5~\fm$ (covering the range of 
typical nuclear scales) indicate that first-order perturbations in $1/a_2$ 
indeed push the state at $B_{4^*}\gsim 8.5~\MeV$ above threshold by about $0.2$ 
and $0.5~\MeV$ for $a_2=20$ and $5~\fm$, respectively.  That corresponds to 
$(B_{4^*}/B_3)^{\NLO(r=0)}$ in the range $0.94\ldots0.98$, compared to 
$(B_{\alpha^*}/B_T)^\text{exp}=0.96$.  The four-body excited state and the 
particle-trimer threshold are close, but both are still far from the 
four-particle threshold; it is reassuring that we can improve the description of 
the excited state at \NLO.

\begin{figure}[tb]
\centering
\includegraphics[width=0.415\textwidth,clip]{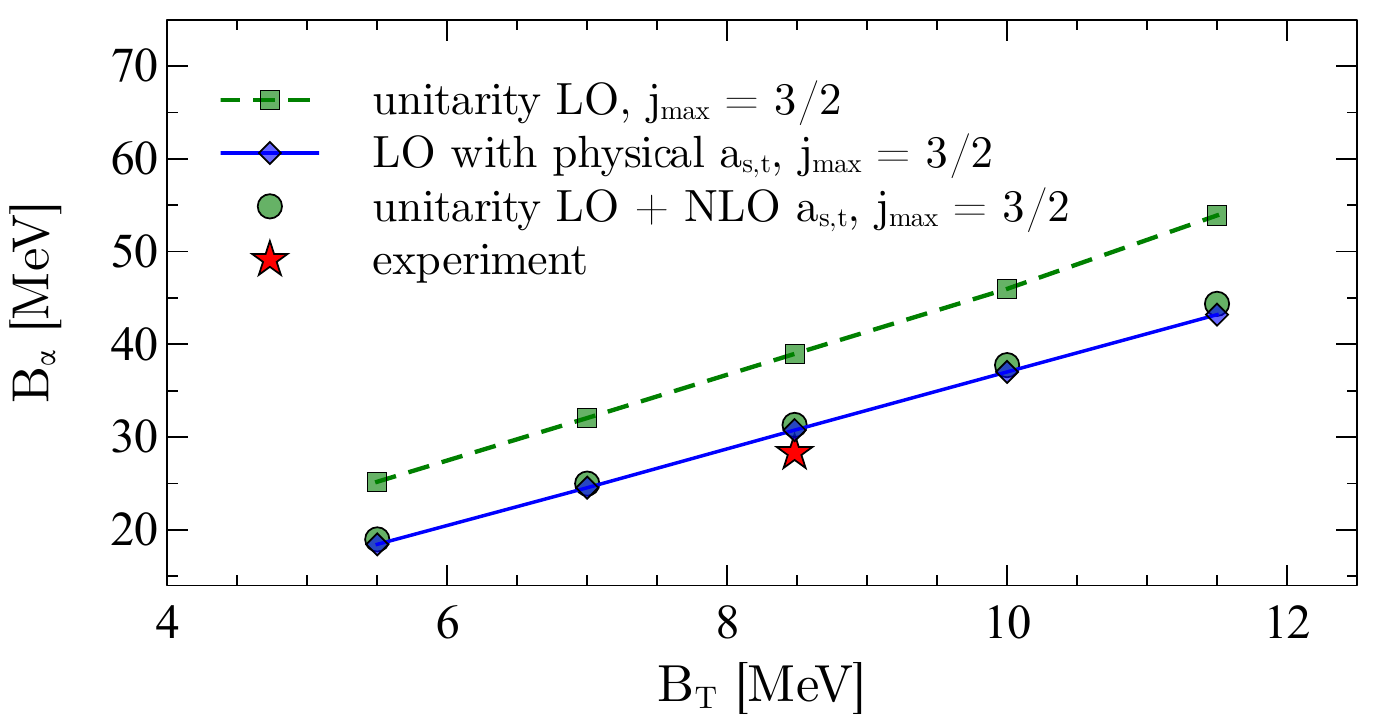}
\caption{Tjon line: correlation between the \FourHe and \ThreeH
  binding energies. (Blue) solid curve: standard pionless LO result; (green)
  dashed upper curve: unitarity limit at LO.  Additional points nearly on top
  of the blue curve: inverse scattering lengths as first-order
  perturbation.  Star: experimental point.}
\label{fig:Tjon-UU}
\end{figure}

As a final test, Fig.~\ref{fig:Tjon-UU} shows the Tjon line, \ie, the
correlation between $3N$ and $4N$ binding energies, obtained by varying
$\Lambda_*$.  All results are calculated with $\jmax=\nicefrac32$ in the $4N$
system and use the $\Lambda \to \infty$ extrapolation discussed above.  The
extrapolation uncertainty is negligible compared to the $30\%$ estimated EFT
truncation error.  The remarkable agreement in Fig.~\ref{fig:En-4He-UU-atad}
persists off the physical point, providing further evidence of the power of a
perturbative expansion around the unitarity limit.  The relation between
triton and \FourHe binding energies is still nearly perfectly linear at \NLO.

Our results suggest good convergence of the expansion around unitarity, both
order by order and to real-world values.  While the condition $Q_AR<1$ is
better satisfied for lighter systems, it may provide at least qualitative
insight into the binding mechanisms of even the heaviest nuclei
($B_A/A\approx8\;\MeV$).  Indeed, the rate of convergence in observables
provides ample evidence that $Q_3R$ is much smaller than its \apriori estimate,
see \eg~Refs.~\cite{Bedaque:2002yg,Griesshammer:2004pe,
Sadeghi:2006fc,Griesshammer:2011md,Vanasse:2013sda,Vanasse:2014kxa,
Konig:2015aka}, and is suggestive that $Q_4R$ is smaller,
too~\cite{Kirscher:2009aj,Kirscher:2015ana,Lensky:2016djr}.  There is also
circumstantial evidence that this may hold for $A>4$~\cite{Bazak:2016wxm}.

A recent expansion around the $SU(4)_W$ limit with the averaged physical
values of the ERE parameters at \LO~found good convergence for $3N$ binding
energies and radii~\cite{Vanasse:2016umz}.  Together with the fact that 
$SU(4)_W$ has some success in heavier nuclei (see 
\eg~Refs.~\cite{DeForest:1965,Lutostansky:2015lza} and references therein),
this adds further credibility to our conjecture.  In the future, we will
investigate our expansion in heavier nuclei, such as the isoscalars
${}^{16}$O~\cite{Contessi:2017rww} and \SixLi~\cite{Stetcu:2006ey}, including
observables beyond binding energies, for example electromagnetic form
factors.  For nuclear matter, saturation energies and densities are correlated
(Coester line)~\cite{Coester:1970ai} and conjectured to be correlated to the 
$3N$ binding energy~\cite{Delfino:2007zu}.  In the unitarity limit, this can be 
understood from discrete scale invariance~\cite{vanKolck:2017}.

In summary, we demonstrated that the physics of $A\le 4$ nucleons is governed
to a good first approximation by a single parameter $\Lambda_*$, with
controlled corrections stemming from deviations from unitarity, the interaction 
range, and isospin-breaking effects.  We conjecture that it also
converges for other light nuclei and speculate about its relevance for heavy
nuclei and nuclear matter.  It may not be a surprise that our results in the
unitarity limit are perturbatively close to those where the physical
scattering lengths are used at \LO.  Surprising is, however, how well the
expansion appears to converge.  Our expansion turns the focus away from details 
of the two-body system, which has traditionally been taken as a starting point 
to the structure of higher-body bound states, and shifts it to a three-body 
interaction that underlies systems around 
unitarity~\cite{Braaten:2004rn,vonStecher:2010,Bazak:2016wxm}.  That adds the 
intriguing possibility that the approach developed here for nucleons may prove 
successful also in atomic and molecular physics, where finite scattering lengths 
are currently treated nonperturbatively.

\begin{acknowledgments}
  We thank J.~Carbonell, R.~Furnstahl, A.~Gezerlis and D.~R.~Phillips for
  discussions; the Institute for Nuclear Theory at the University of
  Washington for hospitality during program INT 16-1 \emph{Nuclear Physics
  from Lattice QCD}; and the participants of the EMMI RRTF workshop
  \emph{ER15-02: Systematic Treatment of the Coulomb Interaction in Few-Body
  Systems} at GSI as well as of the workshop \emph{Lattice Nuclei, Nuclear
  Physics and QCD - Bridging the Gap} at the ECT*.  This work is supported
  in part by the NSF under award PHY--1306250 (SK); by the DOE, NUCLEI SciDAC
  Collaboration award DE-SC0008533 (SK); by the DOE, Office of Science, Office
  of Nuclear Physics, under award numbers DE-SC0015393 (HWG) and
  DE-FG02-04ER41338 (UvK); by the BMBF under contract 05P15RDFN1 (HWH); by the
  DFG through SFB 1245 (HWH); by the Dean's Research Chair programme of the
  Columbian College of Arts and Sciences of The George Washington University
  (HWG); as well as by the ERC Grant No.\ 307986 STRONGINT (SK).  Some results
  reported here were obtained using computing time made available by the State
  of Hesse on the Lichtenberg High-Performance Computer at TU Darmstadt.
\end{acknowledgments}

\end{document}